\documentclass[conference]{IEEEtran}
\usepackage{amsfonts}
\usepackage{mathrsfs}
\usepackage{amssymb,amsmath}

\usepackage{algorithm}
\usepackage{algorithmic}
\usepackage{amsmath,amssymb,epsfig,graphics,subfigure}
\usepackage{theorem}
\usepackage{array,color}
\usepackage[compress]{cite}

\theoremheaderfont{\normalfont\bfseries}

\begin{document}

\title{Joint Robust Weighted LMMSE Transceiver Design for Dual-Hop AF Multiple-Antenna Relay Systems}

\author{Chengwen Xing$^*$, Shaodan Ma$^\dagger$, Zesong Fei$^*$, Yik-Chung Wu$^\dagger$ and Jingming Kuang$^*$
   \\ $^*$School of Information and Electronics, Beijing Institute of Technology, Beijing, China
\\Email: \{xingchengwen\}@gmail.com \ \{feizesong, jmkuang\}@bit.edu.cn
   \\ $^\dagger$Department of Electrical and Electronic Engineering, The University of Hong Kong, Hong Kong
    \\ Email: \{sdma, ycwu\}@eee.hku.hk
}

\maketitle

\begin{abstract}

In this paper, joint transceiver design for dual-hop amplify-and-forward (AF) MIMO relay systems with Gaussian distributed channel estimation errors in both two hops is investigated. Due to the fact that various linear transceiver designs can be transformed to a weighted linear minimum mean-square-error (LMMSE) transceiver design with specific weighting matrices, weighted mean square error (MSE) is chosen as the performance metric. Precoder matrix at source, forwarding matrix at relay and equalizer matrix at destination are jointly designed with channel estimation errors taken care of by Bayesian philosophy. Several existing algorithms are found to be special cases of the proposed solution. The performance advantage of the proposed robust design is demonstrated by the simulation results.

%\textit{Keywords}:  Amplify-and-forward (AF), precoder, equalizer,
%forwading matrix, multiple-input multiple-output (MIMO), minimum
%mean-square-error (MMSE), relay.

\end{abstract}

\section{Introduction}
\label{sect:intro}

 \IEEEPARstart{B}y deploying relays, cooperative communication has great potential to improve system performance \cite{Laneman04}. Among various relaying strategies, amplify-and-forward (AF) scheme is attractive for practical implementation due to its low complexity. On other hand, it is well-established that employing multiple antennas is beneficial to improve wireless system performance due to spatial diversity and multiplexing gains. Consequently, in order to obtain the virtues of these two techniques, combination of AF transmission and multi-input multi-output (MIMO) systems attracts more and more researchers' interest.

Linear transceiver design for AF MIMO relay systems has been widely researched in \cite{Guan08,Mo09,Rong09,Tseng09,Li09}. Unfortunately, most of the existing works assume channel state information (CSI) is perfectly known. However channel estimation errors are inevitable in practical systems.  Robust design, which takes channel estimation errors into account, is therefore desirable in practice. Joint robust design of relay forwarding matrix and destination equalizer under channel estimation errors has been considered in \cite{Xing10}. However in \cite{Xing10}, source precoder design, which provides further degree of freedom for improving system performance, is not considered. In this paper, we take a step further to jointly design source precoder matrix, relay forwarding matrix and destination equalizer matrix under channel estimation errors.\let\oldthefootnote\thefootnote
\renewcommand\thefootnote{}
\footnote{This research work was supported in part by Sino-Swedish IMT-Advanced and Beyond Cooperative Program under Grant No.2008DFA11780 and Program for Changjiang Scholars and Innovative Research Team in University.}
\let\thefootnote\oldthefootnote

There are two main kinds of criteria for transceiver design: capacity maximization and mean-square-error (MSE) minimization.
In fact, both kinds of criteria can be formulated as weighted MSE minimization problems \cite{Sampth01}. Therefore, weighted MSE is chosen as the objective function in this paper.

In this paper, the channel estimation errors are assumed to be Gaussian distributed, which is generally true when linear channel estimation algorithms are adopted \cite{Kay93}. The structures of the optimal solutions for the robust joint transceiver design with minimum weighted MSE as the objective are first derived. Based on these optimal structures, the transceiver design is then significantly simplified and iterative water-filling is adopted to solve the problem. Finally, the performance gain of the proposed robust design is verified by simulation.

The following notations are used throughout this paper. Boldface
lowercase letters denote vectors, while boldface uppercase letters
denote matrices. The notations  ${\bf{Z}}^{\rm{T}}$, ${\bf{Z}}^{*}$ and ${\bf{Z}}^{\rm{H}}$ denote the transpose, conjugate and
conjugate transpose of the matrix ${\bf{Z}}$, respectively and ${\rm{Tr}}({\bf{Z}})$ is the
trace of the matrix ${\bf{Z}}$. The symbol ${\bf{I}}_{M}$ denotes an
$M \times M$ identity matrix, while ${\bf{0}}_{M,N}$ denotes an $M
\times N$ all zero matrix. The notation ${\bf{Z}}^{1/2}$ is the
Hermitian square root of the positive semi-definite matrix
${\bf{Z}}$, such that ${\bf{Z}}^{1/2}{\bf{Z}}^{1/2}={\bf{Z}}$ and
${\bf{Z}}^{1/2}$ is also a Hermitian matrix. The symbol ${\mathbb{E}}$ denotes statistical expectation.

\section{System Model}
\subsection{Transmitted and Received Signals}
In this paper, a three node dual-hop amplify-and-forward (AF) cooperative
communication system is considered. In the considered system, there
is one source with $N_S$ antennas, one relay with $M_{R}$ receive
antennas and $N_R$ transmit antennas, and one destination with $M_D$
antennas. Due to long distance and possibly deep fading, the direct link between the source and destination is not considered in this paper. At the first hop, the source
transmits data to the relay. The received signal, ${\bf{x}}$, at the
relay is
\begin{small}
\begin{align}
{\bf{x}}= {\bf{H}}_{sr}{\bf{P}}{\bf{s}}+{\bf{n}}_1
\end{align}\end{small}where ${\bf{H}}_{sr}$ is the MIMO channel matrix between the source and
the relay, and ${\bf{P}}$ is the precoder matrix at the source. The
vector ${\bf{s}}$ is the $N \times 1$ data vector transmitted by the
source with the covariance matrix
${\bf{R}}_{s}=\mathbb{E}\{{\bf{s}}{\bf{s}}^{\rm{H}}\}={\bf{I}}_N$.
Furthermore, ${\bf{n}}_1$ is the additive Gaussian noise vector with
correlation matrix ${\bf{R}}_{n_1}=\sigma_{n_1}^2{\bf{I}}_{M_R}$.

At the relay, the received signal ${\bf{x}}$ is multiplied by a
forwarding matrix ${\bf{F}}$. Then the resultant signal is
transmitted to the destination. The received signal ${\bf{y}}$ at
the destination can be written as
\begin{small}
\begin{equation}
\label{equ:signal} {\bf{y}} = {{\bf{H}}_{rd} {\bf{F}}
{\bf{H}}_{sr}{\bf{P}}{\bf{s}}}  + {{\bf{H}}_{rd} {\bf{F}}{\bf{n}}_1
} + {\bf{n}}_2,
\end{equation}
\end{small}where ${\bf{H}}_{rd}$ is the MIMO channel matrix between the relay
and the destination, and ${\bf{n}}_2$ is the additive Gaussian noise
vector at the second hop with correlation matrix ${\bf{R}}_{n_2}=\sigma_{n_2}^2{\bf{I}}_{M_D}$.
In order to guarantee the transmitted data ${\bf{s}}$ can be
recovered at the destination, it is assumed that $N_S$, $M_R$,
$N_R$, and $M_D$ are greater than or equal to $N$ \cite{Guan08}.

\subsection{Channel Estimation Error Model}
Channel state information is usually estimated via using training sequences. In a dual-hop system, there are two channels to be estimated. For implementation simplicity, training sequences are transmitted from both source and destination to the relay. At the relay, the two channels are separately estimated and then based on the channel estimates, transceivers are jointly designed. After that, the designed transceiver matrices are forward to the source and destination, respectively. In general, two hop channels can
be written as
\begin{small}
\begin{align}
& {\bf{H}}_{sr}={\bf{\bar H}}_{sr}+\underbrace{{\boldsymbol{\Sigma}}_{sr}^{{1/2}}{\bf{H}}_{W,sr}{\boldsymbol{\Psi}}_{sr}^{1/2}}_{\triangleq\Delta{\bf{H}}_{sr}},  \ \ {\bf{H}}_{rd}={\bf{\bar H}}_{rd}+\underbrace{{\boldsymbol{\Sigma}}_{rd}^{{1/2}}{\bf{H}}_{W,rd}{\boldsymbol{\Psi}}_{rd}^{1/2}}_
{\triangleq \Delta{\bf{H}}_{rd}},
\end{align}\end{small}where ${\bf{\bar H}}_{sr}$ and ${\bf{\bar H}}_{rd}$ are the estimated channels, and $\Delta{\bf{H}}_{sr}$ and $\Delta{\bf{H}}_{rd}$
are the corresponding channel estimation errors whose elements are
zero mean Gaussian random variables. Furthermore, the $M_R \times
N_S$ matrix $\Delta{\bf{H}}_{sr}$ can be decomposed as
$\Delta{\bf{H}}_{sr}={\boldsymbol{\Sigma}}_{sr}^{{1/2}}{\bf{H}}_{W,sr}{\boldsymbol{\Psi}}_{sr}^{1/2}$, where the elements of the $M_R \times N_S$ matrix
${\bf{H}}_{W,sr}$ are independent and identically distributed
(i.i.d.) Gaussian random variables with zero mean and unit variance. When MMSE channel estimator is used, the row
and column covariance matrices of $\Delta{\bf{H}}_{sr}$ can be derived as \cite{Kay93}
\begin{small}
\begin{align}
\label{R_1}
{\boldsymbol{\Sigma}}_{sr}={\bf{I}}_{M_R},  \ \ \ \ {\boldsymbol{\Psi}}_{sr}^{\rm{T}}=(\sigma_{h_{sr}}^{-2}{\bf{I}}_{N_S}+\frac{1}{\sigma_{n_1}^2}{\bf{D}_1}^{*}{\bf{D}}_1^{\rm{T}})^{-1}
\end{align}\end{small}where $\sigma_{h_{sr}}^2$ is channel variance in the first hop and ${\bf{D}}_1$ is the training sequence used in the first hop. Similarly, the row
and column covariance matrices of $\Delta{\bf{H}}_{rd}$, can also be derived to be
\begin{small}
\begin{align}
\label{R_2}
{\boldsymbol{\Sigma}}_{rd}=(\sigma_{h_{rd}}^{-2}{\bf{I}}_{M_D}
+\frac{1}{\sigma_{n_2}^2}{\bf{D}}_2{\bf{D}}_2^{\rm{H}})^{-1},
\ \ \ \ {\boldsymbol{\Psi}}_{rd}^{\rm{T}}={\bf{I}}_{N_R}
\end{align}\end{small}where $\sigma_{h_{rd}}^2$ denotes the channel variance of the second hop and  ${\bf{D}}_2$ represents the training sequence used in the second hop. Notice that the directions of training sequences transmitted in the two hops are opposite. Therefore, the results given by (\ref{R_1}) and ({\ref{R_2}) are different.

\section{Problem Formulation}

At the destination, a linear equalizer ${\bf{G}}$ is adopted to
detect the data vector ${\bf{s}}$. The mean-square-error (MSE)
matrix is $\mathbb{E}\{({\bf{G}}{\bf{y}}-{\bf{s}})({\bf{G}}{\bf{y}}-{\bf{s}})^{\rm{H}}\} $ , where the expectation is taken with respect to random data, channel estimation errors, and noise.  In \cite{Xing10}, it is shown that
\begin{small}
\begin{align}
\label{MSE_final} &\mathbb{E}\{({\bf{G}}{\bf{y}}-{\bf{s}})({\bf{G}}{\bf{y}}-{\bf{s}})^{\rm{H}}\}\nonumber \\& ={\bf{G}}({\bf{\bar
H}}_{rd}{\bf{F}}{\bf{R}}_{\bf{x}}{\bf{F}}^{\rm{H}} {\bf{\bar
H}}_{rd}^{\rm{H}}+{\bf{K}}_2){\bf{G}}^{\rm{H}}+ {\bf{I}}_N-(
{\bf{P}}^{\rm{H}}{\bf{\bar
H}}_{sr}^{\rm{H}}{\bf{F}}^{\rm{H}}{\bf{\bar
H}}_{rd}^{\rm{H}}{\bf{G}}^{\rm{H}})\nonumber \\
&- (
{\bf{G}}{\bf{\bar H}}_{rd}{\bf{F}}{\bf{\bar H}}_{sr}{\bf{P}}
),
\end{align}\end{small}where matrices ${\bf{R}}_{\bf{x}}$ and ${\bf{K}}_2$ are defined as\begin{small}\begin{align}
\label{R_x}
{\bf{R}}_{\bf{x}}&=\mathbb{E}\{{\bf{x}}{\bf{x}}^{\rm{H}}\}={\bf{\bar H}}_{sr}{\bf{P}}{\bf{P}}^{\rm{H}}{\bf{\bar H}}_{sr}^{\rm{H}}+{\bf{K}}_1 \nonumber \\{\bf{K}}_1&={\rm{Tr}}({\bf{P}}{\bf{P}}^{\rm{H}}{\boldsymbol
\Psi}_{sr}){\boldsymbol \Sigma}_{sr}+{\bf{R}}_{n_1}\nonumber \\
{\bf{K}}_2&={\rm{Tr}}({\bf{F}}{\bf{R}}_{\bf{x}} {\bf{F}}^{\rm{H}}
{\boldsymbol\Psi} _{rd}
   ){\boldsymbol\Sigma} _{rd}+{\bf{R}}_{n_2}.
\end{align}\end{small}It is obvious that ${\bf{R}}_{\bf{x}}$ is the covariance matrix of the received signal at the relay.
Based on the MSE matrix given by (\ref{MSE_final}), a more general performance metric for transceiver design is the weighted MSE \cite{Sampth01}:
\begin{small}\begin{align}
{\rm{MSE}}_W({\bf{G}},{\bf{F}},{\bf{P}})&={{\mathbb{E}}\{({\bf{G}}{\bf{y}}-{\bf{s}})^{\rm{H}} {\bf{W}}({\bf{G}}{\bf{y}}-{\bf{s}})\}}\nonumber \\
&={\rm{Tr}}[{\bf{W}}{{\mathbb{E}}\{({\bf{G}}{\bf{y}}-{\bf{s}}) ({\bf{G}}{\bf{y}}-{\bf{s}})^{\rm{H}}\}}],
\end{align}\end{small}where ${\bf{W}}$ is a $N \times N$ positive semi-definite weighting matrix. Under transmit power constraints at the source and relay, the
optimization problem of transceiver design is formulated as
\begin{small}\begin{align}
\label{prob:opt}
{\rm{P1:}} \quad
& \min_{{\bf{G}},{\bf{F}},{\bf{P}}} \ \ {\rm{MSE}}_W({\bf{G}},{\bf{F}},{\bf{P}}) \nonumber \\
& \ {\rm{s.t.}} \ \ \ \ {\rm{Tr}}({\bf{P}}{\bf{P}}^{\rm{H}}) \le P_{s}, \ \
{\rm{Tr}}({\bf{F}}{\bf{R}}_{\bf{x}}{\bf{F}}^{\rm{H}}) \le P_{r}.
\end{align}\end{small}

Notice that based on the definition of ${\bf{R}}_{\bf{x}}$ in (\ref{R_x}), ${\bf{R}}_{{\bf{x}}}$ is a function of ${\bf{P}}$. In order to simplify the analysis, we define a new variable\begin{small}\begin{align}
\label{Tilde_F}
{\bf{\tilde F}}\triangleq
{\bf{F}}{\bf{K}}_1^{1/2}(\underbrace{{\bf{K}}_1^{-1/2}{\bf{\bar
H}}_{sr}{\bf{P}}{\bf{P}}^{\rm{H}}{\bf{\bar
H}}_{sr}^{\rm{H}}{\bf{K}}_1^{-1/2}+{\bf{I}}_{M_R}}_{\triangleq {\boldsymbol \Pi}_{\bf{P}}})^{1/2},
\end{align}\end{small}based on which ${\rm{Tr}}({\bf{F}}{\bf{R}}_{\bf{x}}{\bf{F}}^{\rm{H}})={\rm{Tr}}({\bf{\tilde F}}{\bf{\tilde F}}^{\rm{H}})$ and the two constraints in the optimization problem (\ref{prob:opt}) becomes independent. Furthermore, with (\ref{Tilde_F}) the objective function becomes\begin{small}\begin{align}
\label{MSE}
&{\rm{MSE}}_W({\bf{G}},{\bf{\tilde F}},{\bf{P}}) \nonumber \\
 =& {\rm{Tr}}[{\bf{W}}{\bf{G}}({\bf{\bar H}}_{rd}{\bf{\tilde
F}}{\bf{\tilde F}}^{\rm{H}} {\bf{\bar
H}}_{rd}^{\rm{H}}+{\bf{K}}_2){\bf{G}}^{\rm{H}}]\nonumber
\\
&-{\rm{Tr}}({\bf{W}}{\bf{G}}{\bf{\bar H}}_{rd}{\bf{\tilde
F}}{\boldsymbol \Pi}_{\bf{P}}^{-1/2}{\bf{K}}_1^{-1/2}{\bf{\bar
H}}_{sr}{\bf{P}})\nonumber \\
& -{\rm{Tr}}({\bf{W}}{\bf{P}}^{\rm{H}}{\bf{\bar
H}}_{sr}^{\rm{H}}{\bf{K}}_1^{-1/2}{\boldsymbol \Pi}_{\bf{P}}^{-1/2}{\bf{\tilde
F}}^{\rm{H}}{\bf{\bar H}}_{rd}^{\rm{H}}{\bf{G}}^{\rm{H}})+{\rm{Tr}}(
{\bf{W}}).
\end{align}\end{small}As in the optimization problem (\ref{prob:opt}) there is no constraint on ${\bf{G}}$, the optimal solution for ${\bf{G}}$ can therefore be written as a function of ${\bf{P}}$ and ${\bf{\tilde F}}$ \cite{Xing10}, and is given by
\begin{small}
\begin{align}
\label{G} & {\bf{G}}=({\bf{\bar H}}_{rd}{\bf{\tilde
F}}{\boldsymbol \Pi}_{\bf{P}}^{-1/2}{\bf{K}}_1^{-1/2}{\bf{\bar
H}}_{sr}{\bf{P}})^{\rm{H}}({\bf{\bar H}}_{rd}{\bf{\tilde
F}}{\bf{\tilde F}}^{\rm{H}}{\bf{\bar
H}}_{rd}^{\rm{H}}+{\bf{K}}_2)^{-1}.
\end{align}\end{small}Substituting (\ref{G}) into the MSE formulation
(\ref{MSE}), the weighted MSE can be rewritten as
\begin{small}
\begin{align}
\label{MSE_Matrix}
&{\rm{\overline{MSE}}}_W({\bf{\tilde
F}},{\bf{P}})\nonumber \\ & ={\rm{Tr}}({\bf{W}})-{\rm{Tr}}[({\bf{\bar
H}}_{rd}{\bf{\tilde
F}}{\boldsymbol \Pi}_{\bf{P}}^{-1/2}{\bf{K}}_1^{-1/2}{\bf{\bar
H}}_{sr}{\bf{P}}{\bf{W}}^{1/2})^{\rm{H}} \nonumber \\
& \times({\bf{\bar H}}_{rd}{\bf{\tilde
F}}{\bf{\tilde F}}^{\rm{H}} {\bf{\bar
H}}_{rd}^{\rm{H}}+{\bf{K}}_2)^{-1}({\bf{\bar H}}_{rd}{\bf{\tilde
F}}{\boldsymbol \Pi}_{\bf{P}}^{-1/2}{\bf{K}}_1^{-1/2}{\bf{\bar H}}_{sr}{\bf{P}}{\bf{W}}^{1/2})]
.
\end{align}\end{small}Finally, it can be easily proved that the optimal ${\bf{P}}$ and ${\bf{\tilde
F}}$ must occur on the boundary \cite{Xing10}
\begin{small}
\begin{align}
{\rm{Tr}}({\bf{P}}{\bf{P}}^{\rm{H}})=P_s, \ \ {\rm{Tr}}({\bf{\tilde
F}}{\bf{\tilde F}}^{\rm{H}})=P_r.
\end{align}\end{small}Clearly, it means that for the minimum weighted MSE, both relay and source should transmit at the maximum power. As a result, the optimization problem for joint transceiver design is formulated as
\begin{small}
\begin{align}
\label{prob:1}
{\rm{P2:}} \quad
& \min_{{\bf{\tilde F}},{\bf{P}}} \ \ {\rm{\overline{MSE}}}_W({\bf{\tilde
 F}},{\bf{P}}) \nonumber \\
& \ {\rm{s.t.}} \ \ \ {\rm{Tr}}({\bf{\tilde F}}{\bf{\tilde
F}}^{\rm{H}}) = P_{r}, \ \  {\rm{Tr}}({\bf{P}}{\bf{P}}^{\rm{H}})
= P_{s}.
\end{align}\end{small}In the following, the structures of the optimal solutions are first derived and then the transceiver design problem is simplified and solved in Section~\ref{Sect:solution}.

\section{The structure of optimal ${\bf{\tilde F}}$}

With\begin{small}\begin{align}
\label{A_M_Define}
&{\bf{A}}\triangleq ({\boldsymbol{\Pi}}_{\bf{P}}^{-1/2}{\bf{K}}_1^{-1/2}{\bf{\bar
H}}_{sr}{\bf{P}}{\bf{W}}^{1/2})\nonumber \\
&   {\bf{M}}\triangleq {\bf{\tilde
F}}^{\rm{H}}{\bf{\bar H}}_{rd}^{\rm{H}}({\bf{\bar
H}}_{rd}{\bf{\tilde F}}{\bf{\tilde F}}^{\rm{H}} {\bf{\bar
H}}_{rd}^{\rm{H}}+{\bf{K}}_2)^{-1}{\bf{\bar H}}_{rd}{\bf{\tilde F}},
\end{align}\end{small}the weighted MSE can be further reformulated as
\begin{small}
\begin{align}
\label{MSE_simple}
{\rm{\overline{MSE}}}_{\bf{W}}({\bf{\tilde
 F}},{\bf{P}})={\rm{Tr}}({\bf{W}})-{\rm{Tr}}({\bf{A}}^{\rm{H}}{\bf{M}}{\bf{A}}).
 \end{align}\end{small}Using the matrix inversion lemma, \begin{small}${\bf{M}}={\bf{I}}-({\bf{\tilde
F}}^{\rm{H}}{\bf{\bar H}}_{rd}^{\rm{H}}
{\bf{K}}_2^{-1}{\bf{\bar H}}_{rd}{\bf{\tilde F}}+{\bf{I}})^{-1}$\end{small}, the weighted MSE further becomes
\begin{small}
\begin{align}
\label{weighted_MSE}
{\rm{\overline{MSE}}}_{\bf{W}}({\bf{\tilde
 F}},{\bf{P}})&={\rm{Tr}}({\bf{W}})-{\rm{Tr}}({\bf{A}}^{\rm{H}}{\bf{A}})\nonumber
\\ & +{\rm{Tr}}[{\bf{A}}^{\rm{H}}({\bf{\tilde F}}^{\rm{H}}{\bf{\bar
H}}_{rd}^{\rm{H}}{\bf{K}}_2^{-1}{\bf{\bar
H}}_{rd}{\bf{\tilde F}}+{\bf{I}}_{M_R})^{-1}{\bf{A}}].
\end{align}\end{small}
It should be highlighted that the two constraints in (\ref{prob:1}) are independent and for any given ${\bf{P}}$ the optimization problem for ${\bf{\tilde F}}$ is equivalent to
\begin{small}
\begin{align}
\label{Prob:F_Final} &{\min_{{\bf{\tilde F}}}} \ \ \
{\rm{Tr}}[{\bf{A}}^{\rm{H}}({\bf{\tilde
F}}^{\rm{H}}{\bf{\bar H}}_{rd}^{\rm{H}}
{\bf{K}}_2^{-1}{\bf{\bar H}}_{rd}{\bf{\tilde F}}+{\bf{I}}_{M_D})^{-1}{\bf{A}}] \nonumber \\
& \ {\rm{s.t.}} \ \ \
 {\rm{Tr}}({\bf{\tilde F}}{\bf{\tilde F}}^{\rm{H}})= P_{r}
\end{align}\end{small}where the constant parts independent of ${\bf{\tilde F}}$ have been neglected. With the fact that ${\boldsymbol \Psi}_{rd}={\bf{I}}$ from (\ref{R_2}), we have
\begin{small}
\begin{align}
{\bf{K}}_2={\rm{Tr}}({\bf{\tilde F}}{\bf{\tilde F}}^{\rm{H}}{\boldsymbol \Psi}_{rd}){\boldsymbol \Sigma}_{rd}+{\sigma_{n_2}^2}{\bf{I}}_{M_D}=P_r{\boldsymbol \Sigma}_{rd}+{\sigma_{n_2}^2}{\bf{I}}_{M_D}
\end{align}\end{small}which is a constant matrix. Furthermore, based on singular value decomposition
\begin{small}
\begin{align}
\label{A}
{\bf{A}}&={\bf{U}}_{\bf{A}}{\boldsymbol \Lambda}_{\bf{A}}{\bf{V}}_{\bf{A}}^{\rm{H}}, \ \
{\bf{K}}_2^{-1/2}{\bf{\bar H}}_{rd}={\bf{U}}_{rd}{\boldsymbol \Lambda}_{rd}{\bf{V}}_{rd}^{\rm{H}}.
\end{align}\end{small}where the diagonal elements of ${\boldsymbol \Lambda}_{\bf{A}}$ and ${\boldsymbol \Lambda}_{rd}$ are in decreasing order. Using Majorization theory, the objective function of (\ref{Prob:F_Final}) can be transformed into a Schur-concave function of the diagonal elements of \begin{small}$({\bf{U}}_{\bf{A}}^{\rm{H}}{\bf{\tilde
F}}^{\rm{H}}{\bf{\bar H}}_{rd}^{\rm{H}}
{\bf{K}}_2^{-1}{\bf{\bar H}}_{rd}{\bf{\tilde F}}{\bf{U}}_{\bf{A}}+{\bf{I}}_{M_R})^{-1}$\end{small}. Therefore, the optimal solution of (\ref{Prob:F_Final})
has the following structure \cite{Xing10}\begin{small}
\begin{align}
\label{Structure_F}
  {\bf{\tilde F}}={\bf{V}}_{rd,N}{\boldsymbol \Lambda}_{{\bf{\tilde F}}}{\bf{U}}_{{\bf{A}},N}^{\rm{H}}
\end{align}\end{small}where ${\boldsymbol \Lambda}_{{\bf{\tilde F}}}$ is a $N\times N$ diagonal matrix such that the diagonal elements of ${\boldsymbol \Lambda}_{{\bf{\tilde F}}}{\boldsymbol {\tilde \Lambda}}_{rd}^2{\boldsymbol \Lambda}_{{\bf{\tilde F}}}$ are in decreasing order. The diagonal matrix ${\boldsymbol {\tilde \Lambda}}_{rd}$ is the $N\times N$ principal submatrix of ${\boldsymbol {\Lambda}}_{rd}$. The matrices ${\bf{V}}_{rd,N}$ and ${\bf{U}}_{{\bf{A}},N}$ are the first $N$ columns of  ${\bf{V}}_{rd}$ and ${\bf{U}}_{\bf{A}}$, respectively. As ${\bf{A}}$ is a function of ${\bf{P}}$, the value of ${\bf{U}}_{{\bf{A}},N}$ will be given in the next section, after giving the  structure of optimal ${\bf{P}}$.

\section{The structure of optimal ${\bf{P}}$}

Substituting the structure of optimal ${\bf{\tilde F}}$ (\ref{Structure_F}) into the definition of ${\bf{M}}$ in (\ref{A_M_Define}), and together with the fact that  the diagonal elements of ${\boldsymbol \Lambda}_{{\bf{\tilde F}}}{\boldsymbol {\tilde \Lambda}}_{rd}^2{\boldsymbol \Lambda}_{{\bf{\tilde F}}}$ are in decreasing order, we have
 \begin{small}
 \begin{align}
 \label{M}
 {\bf{M}}={\bf{U}}_{{\bf{A}},N}\underbrace{[{\bf{I}}_N-({\boldsymbol {\Lambda}}_{{\bf {\tilde F}}}{\boldsymbol {\tilde \Lambda}}_{rd}^2{\boldsymbol {\Lambda}}_{{\bf {\tilde F}}}+{\bf{I}}_N)^{-1}]}_{\triangleq {\boldsymbol {\tilde \Lambda}}_{\bf{M}}}
 {\bf{U}}_{{\bf{A}},N}^{\rm{H}}
 \end{align}\end{small}where ${\boldsymbol {\tilde \Lambda}}_{{\bf{ M}}}$ is a $N\times N$ diagonal matrix with diagonal elements in decreasing order. It is also straightforward that (\ref{M}) is the eigen-decomposition of ${\bf{ M}}$. Based on (\ref{M}), and the singular value decomposition of ${\bf{A}}$ given in (\ref{A}), after a straightforward substitution we have the following identity
 \begin{small}
 \begin{align}
 \label{AMA}
 {\bf{A}}^{\rm{H}}{\bf{M}}{\bf{A}}={\bf{V}}_{\bf{A}}{\boldsymbol{\tilde
\Lambda}}_{{\bf{A}}}{\boldsymbol{\tilde
\Lambda}}_{{\bf{M}}}{\boldsymbol{\tilde
\Lambda}}_{{\bf{A}}}{\bf{V}}_{\bf{A}}^{\rm{H}}={\bf{V}}_{\bf{A}}{\boldsymbol{\tilde
\Lambda}}_{{\bf{M}}}{\bf{V}}_{\bf{A}}^{\rm{H}}
{\bf{A}}^{\rm{H}}{\bf{A}}
 \end{align}
 \end{small}where ${\boldsymbol{\tilde
\Lambda}}_{{\bf{A}}}$ is the $N\times N$ principal submatrix of ${\boldsymbol{
\Lambda}}_{{\bf{A}}}$ and is a diagonal matrix with diagonal elements in decreasing order. Using (\ref{AMA}), the weighted MSE given by (\ref{MSE_simple}) is rewritten as
\begin{small}
\begin{align}
\label{equ_A_M} {\rm{\overline{MSE}}}_W({\bf{\tilde
 F}},{\bf{P}})&={\rm{Tr}}({\bf{W}})-{\rm{Tr}}({\bf{A}}^{\rm{H}}{\bf{M}}{\bf{A}}) \nonumber \\
 &={\rm{Tr}}({\bf{W}})-{\rm{Tr}}({\bf{V}}_{\bf{A}}{\boldsymbol{\tilde
\Lambda}}_{{\bf{M}}}{\bf{V}}_{\bf{A}}^{\rm{H}}
{\bf{A}}^{\rm{H}}{\bf{A}}).
\end{align}\end{small}It is shown in Appendix A that for the minimum MSE the following identity holds
\begin{small}
\begin{align}
\label{V_A}
{\bf{V}}_{\bf{A}}={\bf{U}}_{\bf{W}},
\end{align}\end{small}where the unitary matrix ${\bf{U}}_{\bf{W}}$ is obtained from the eigen-decomposition of ${\bf{W}}={\bf{U}}_{\bf{W}}{\boldsymbol \Lambda}_{\bf{W}}{\bf{U}}_{\bf{W}}^{\rm{H}}$ in which ${\boldsymbol \Lambda}_{\bf{W}}$ is a diagonal matrix with decreasing diagonal elements. The above equation will also be useful in the following derivation.

From (\ref{V_A}) and the definition of ${\bf{A}}$ in
(\ref{A_M_Define}) and using the matrix inversion lemma again, the weighted MSE (\ref{equ_A_M}) can be further
rewritten as
\begin{small}
\begin{align}
\label{MSE_F}
& {\rm{\overline{MSE}}}_W({\bf{\tilde F}},{\bf{P}}) \nonumber \\
=&{\rm{Tr}}({\bf{W}})-{\rm{Tr}}[({\boldsymbol{\Pi}}_{\bf{P}}^{-1/2}{\bf{K}}_1^{-1/2}{\bf{\bar
H}}_{sr}{\bf{P}}){\bf{W}}^{1/2}{\bf{U}}_{{\bf{W}}}{\boldsymbol{\tilde
\Lambda}}_{{\bf{M}}}{\bf{U}}_{{\bf{W}}}^{\rm{H}}{\bf{W}}^{1/2}\nonumber \\
& \times(
{\boldsymbol{\Pi}}_{\bf{P}}^{-1/2}{\bf{K}}_1^{-1/2}{\bf{\bar
H}}_{sr}{\bf{P}})^{\rm{H}}] \nonumber \\
=&{\rm{Tr}}({\bf{W}})+{\rm{Tr}}[{\bf{U}}_{{\bf{W}}}{\boldsymbol
\Lambda}_{\bf{W}}^{1/2}{\boldsymbol {\tilde
\Lambda}}_{{\bf{M}}}{\boldsymbol {
\Lambda}}_{\bf{W}}^{1/2}{\bf{U}}_{{\bf{W}}}^{\rm{H}}({\bf{P}}^{\rm{H}}{\bf{\bar
H}}_{sr}^{\rm{H}} {\bf{K}}_1^{-1}{\bf{\bar H}}_{sr}
{\bf{P}}+{\bf{I}})^{-1}] \nonumber
\\& -{\rm{Tr}}({\bf{W}}^{1/2}{\bf{U}}_{{\bf{W}}}{\boldsymbol{\tilde
\Lambda}}_{{\bf{M}}}{\bf{U}}_{{\bf{W}}}^{\rm{H}}{\bf{W}}^{1/2}).
\end{align}\end{small}Notice that based on the definition of ${\bf{M}}$ in
(\ref{A_M_Define}), ${\boldsymbol{\tilde \Lambda}}_{\bf{M}}$ is a function
of ${\bf{\tilde F}}$ only and independent of ${\bf{P}}$. Only the second term in (\ref{MSE_F}) is a function of ${\bf{P}}$. Then, the optimization
problem for ${\bf{P}}$ becomes
\begin{small}
\begin{align}
\label{Prob:P_Final}
 &{\min_{{\bf{p}}}} \ \ \
{\rm{Tr}}[{\bf{U}}_{{\bf{W}}}{\boldsymbol
\Lambda}_{\bf{W}}^{1/2}{\boldsymbol {\tilde
\Lambda}}_{{\bf{M}}}{\boldsymbol
\Lambda}_{\bf{W}}^{1/2}{\bf{U}}_{{\bf{W}}}^{\rm{H}} ({\bf{P}}^{\rm{H}}{\bf{\bar
H}}_{sr}^{\rm{H}}
{\bf{K}}_1^{-1}{\bf{\bar H}}_{sr} {\bf{P}}+{\bf{I}}_N)^{-1}]\nonumber \\
& \ {\rm{s.t.}} \ \ \
 {\rm{Tr}}({\bf{P}}{\bf{P}}^{\rm{H}})= P_{s}.
\end{align}\end{small}
For source precoder design, the main difference from forwarding matrix ${\bf{\tilde F}}$ design is that ${\bf{K}}_1$ is not constant. As mentioned previously ${\boldsymbol \Sigma}_{sr}={\bf{I}}$ (\ref{R_1}), then ${\bf{K}}_1$ equals to
\begin{small}
\begin{align}
\label{eta_f}
{\bf{K}}_1={[{\rm{Tr}}({\bf{P}}{\bf{P}}^{\rm{H}}{\boldsymbol \Psi}_{sr})+\sigma_{n_1}^2]}{\bf{I}}_{M_R}{\triangleq {\eta}_p{\bf{I}}}_{M_R}.
\end{align}\end{small}With the power constraint ${\rm{Tr}}({\bf{P}}{\bf{P}}^{\rm{H}})=P_s$, we have
\begin{small}
\begin{align}
\label{P_eta}
\eta_{p}&={\rm{Tr}}({\bf{P}}{\bf{P}}^{\rm{H}}{\boldsymbol \Psi}_{sr})+\sigma_{n_1}^2 \nonumber \\ &={\rm{Tr}}({\bf{P}}{\bf{P}}^{\rm{H}}{\boldsymbol \Psi}_{sr})+\sigma_{n_1}^2
\underbrace{{\rm{Tr}}({\bf{P}}{\bf{P}}^{\rm{H}})/P_s}_{=1} \nonumber \\ &={\rm{Tr}}({\bf{P}}{\bf{P}}^{\rm{H}}(P_{s}{\boldsymbol
\Psi}_{sr}+\sigma_{n_1}^2{\bf{I}}_{N_S}))/P_s.
\end{align}\end{small}From (\ref{P_eta}), the constraint of the optimization problem
(\ref{Prob:P_Final}) becomes as
\begin{small}
\begin{align}
\label{constraint}
&{\rm{Tr}}({\bf{P}}{\bf{P}}^{\rm{H}})={\rm{Tr}}[{\bf{P}}{\bf{P}}^{\rm{H}}(P_{s}{\boldsymbol
\Psi}_{sr}+\sigma_{n_1}^2{\bf{I}}_{N_S})]/\eta_{p}=P_s,
\end{align}\end{small}based on which the optimization problem (\ref{Prob:P_Final}) is equivalent to
\begin{small}
\begin{align}
\label{P_Prob}
 & {\min_{{\bf{P}}}} \ \ \ \
{\rm{Tr}}[{\bf{U}}_{{\bf{W}}}{\boldsymbol
\Lambda}_{\bf{W}}^{1/2}{\boldsymbol {\tilde
\Lambda}}_{{\bf{M}}}{\boldsymbol
\Lambda}_{\bf{W}}^{1/2}{\bf{U}}_{{\bf{W}}}^{\rm{H}}
({\bf{P}}^{\rm{H}}{1}/{\eta_{p}}{\bf{\bar H}}_{sr}^{\rm{H}}{\bf{\bar{H}}}_{sr} {\bf{P}}+{\bf{I}}_N)^{-1}] \nonumber \\
& {\rm{s.t.}} \ \ \ \ \
{\rm{Tr}}[{\bf{P}}{\bf{P}}^{\rm{H}}(P_{s}{\boldsymbol
\Psi}_{sr}+\sigma_{n_1}^2{\bf{I}}_{N_S})]/\eta_{p}= P_{s}.
\end{align}\end{small}

Defining a new variable,
\begin{small}\begin{align}
\label{P_tilde}
{\bf{\tilde P}}=1/\sqrt{\eta_p}(P_{s}{\boldsymbol
\Psi}_{sr}+\sigma_{n_1}^2{\bf{I}}_{N_S})^{1/2}{\bf{P}}
\end{align}\end{small}the optimization problem (\ref{P_Prob}) becomes as (\ref{P_Prob_Final}) as shown at the top of the next page.
\begin{figure*}[!t]
\setcounter{equation}{33}
\begin{small}
\begin{align}
\label{P_Prob_Final}
 & {\min_{{\bf{\tilde P}}}} \ \ \ \
{\rm{Tr}}[{{\bf{U}}_{{\bf{W}}}{\boldsymbol
\Lambda}_{\bf{W}}^{1/2}{\boldsymbol {\tilde
\Lambda}}_{{\bf{M}}}{\boldsymbol
\Lambda}_{\bf{W}}^{1/2}{\bf{U}}_{{\bf{W}}}^{\rm{H}}}
({\bf{\tilde P}}^{\rm{H}}{(P_{s}{\boldsymbol
\Psi}_{sr}+\sigma_{n_1}^2{\bf{I}}_{N_S})^{-1/2}{\bf{\bar H}}_{sr}^{\rm{H}}{\bf{\bar{H}}}_{sr}(P_{s}{\boldsymbol
\Psi}_{sr}+\sigma_{n_1}^2{\bf{I}}_{N_S})^{-1/2}}{\bf{\tilde P}}+{\bf{I}}_N)^{-1}] \nonumber \\
& {\rm{s.t.}} \ \ \ \ \
{\rm{Tr}}({\bf{\tilde P}}{\bf{\tilde P}}^{\rm{H}})= P_{s}
\end{align}\end{small}\hrulefill
\setcounter{equation}{34}
\vspace*{1.5pt}
\end{figure*}This formulation is exactly the same as the optimization problem (\ref{Prob:F_Final}) for ${\bf{\tilde F}}$. Following the same argument for ${\bf{\tilde F}}$ and defining unitary matrices ${\bf{U}}_{sr}$ and ${\bf{V}}_{sr}$ based on the following singular value decomposition
\begin{small}
\begin{align}
{\bf{\bar{H}}}_{sr}(P_{s}{\boldsymbol
\Psi}_{sr}+\sigma_{n_1}^2{\bf{I}}_{N_S})^{-1/2}={\bf{U}}_{sr}{\boldsymbol
\Lambda}_{sr}{\bf{V}}_{sr}^{\rm{H}},
\end{align}\end{small}with the diagonal elements of the diagonal matrix ${\boldsymbol
\Lambda}_{sr}$ in decreasing order, the optimal ${\bf{\tilde P}}$ has the following structure
\begin{small}
\begin{align}
{\bf{\tilde P}}={\bf{V}}_{sr,N}{\boldsymbol \Lambda}_{{\bf{\tilde P}}}{\bf{U}}_{\bf{W}}^{\rm{H}}
\end{align}\end{small}where ${\boldsymbol \Lambda}_{{\bf{\tilde P}}}$ is a $N\times N$ diagonal matrix such that the diagonal elements of ${\boldsymbol \Lambda}_{{\bf{\tilde  P}}}{\boldsymbol {\tilde \Lambda}}_{sr}^2{\boldsymbol \Lambda}_{{\bf{\tilde P}}}$ are in decreasing order. The diagonal matrix ${\boldsymbol {\tilde \Lambda}}_{sr}$ is the $N\times N$ principal submatrix of ${\boldsymbol {\Lambda}}_{sr}$. Furthermore, based on the definition of ${\bf{\tilde P}}$ given by (\ref{P_tilde}), the optimal ${\bf{P}}$ has the following structure
\begin{small}
\begin{align}
{\label{Structure_P}}
 {\bf{P}}&=\sqrt{\eta_p}(P_{s}{\boldsymbol
\Psi}_{sr}+\sigma_{n_1}^2{\bf{I}}_{N_S})^{-1/2}{\bf{V}}_{sr,N}
{\boldsymbol{\Lambda}}_{\bf{\tilde P}}
{\bf{U}}_{{\bf{W}}}^{\rm{H}}.
\end{align}\end{small}Substituting (\ref{Structure_P}) into the definition of ${\bf{A}}$ in (\ref{A_M_Define}), we have ${\bf{U}}_{{\bf{A}},N}={\bf{U}}_{sr,N}$. Therefore, the optimal ${\bf{\tilde F}}$ has the following structure
\begin{small}
\begin{align}
\label{Structure_F_a}
{\bf{\tilde F}}={\bf{V}}_{rd,N}{\boldsymbol \Lambda}_{{\bf{\tilde F}}}{\bf{U}}_{sr,N}^{\rm{H}}.
\end{align}\end{small}\noindent {\textbf{Remark}}: Given (\ref{Structure_P}) and (\ref{Structure_F_a}), the remaining problem is how to determine two diagonal matrices ${\boldsymbol \Lambda}_{\bf{\tilde F}}$ and ${\boldsymbol \Lambda}_{\bf{\tilde P}}$.

Notice that after ${\boldsymbol \Lambda}_{{\bf{\tilde P}}}$ is computed, the remaining unknown parameter in (\ref{Structure_P}) is only $\eta_p$. In order to solve
$\eta_p$, substitute the formulation of ${\bf{P}}$ (\ref{Structure_P}) into the definition of $\eta_p$ in
(\ref{eta_f}), and then we get
\begin{small}
\begin{align}
\eta_p&={\rm{Tr}}({\bf{P}}{\bf{P}}^{\rm{H}}{\boldsymbol
\Psi}_{sr})+\sigma_{n_1}^2 \nonumber \\
&=\eta_p{\rm{Tr}}[{\bf{V}}_{sr,N}^{\rm{H}}
(P_{s}{\boldsymbol
\Psi}_{sr}+\sigma_{n_1}^2{\bf{I}})^{-1/2}{\boldsymbol
\Psi}_{sr}(P_s{\boldsymbol
\Psi}_{sr}+\sigma_{n_1}^2{\bf{I}})^{-1/2}\nonumber \\ &\times
{\bf{V}}_{sr,N}{\boldsymbol{\Lambda}}_{{\bf{\tilde P}}}^2]
+\sigma_{n_1}^2.
\end{align}\end{small}This is a simple linear function of ${\eta}_p$, and $\eta_p$ can be easily solved to be\begin{small}
\begin{align}
\eta_p&={\sigma_{n_1}^2}/\{1-{\rm{Tr}}[{\bf{V}}_{sr,N}^{\rm{H}}(P_{s}{\boldsymbol
\Psi}_{sr}+\sigma_{n_1}^2{\bf{I}})^{-1/2}{\boldsymbol
\Psi}_{sr}\nonumber \\& \times(P_s{\boldsymbol
\Psi}_{sr}+\sigma_{n_1}^2{\bf{I}})^{-1/2}
{\bf{V}}_{sr,N}{\boldsymbol{\Lambda}}_{{\bf{\tilde P}}}^2]\}.
\end{align}\end{small}\section{Proposed Solution for ${\boldsymbol \Lambda}_{\bf{\tilde F}}$ and ${\boldsymbol \Lambda}_{\bf{\tilde P}}$}
\label{Sect:solution}

In this section, the optimal ${\boldsymbol \Lambda}_{\bf{\tilde F}}$ and ${\boldsymbol \Lambda}_{\bf{\tilde P}}$ will be derived. Notice that ${\boldsymbol { \tilde \Lambda}}_{sr}$ and ${\boldsymbol { \tilde \Lambda}}_{rd}$ are the $N\times N$ principal submatrices of ${\boldsymbol {\Lambda}}_{sr}$ and ${\boldsymbol {\Lambda}}_{rd}$, respectively. Based on (\ref{constraint}), substituting the optimal structures (\ref{Structure_P}) and (\ref{Structure_F_a}) into the original optimization problem (\ref{prob:1}), and denoting\begin{small}
\begin{align}
& {\boldsymbol { \tilde \Lambda}}_{sr}={\rm{diag}}\{\lambda_{sr,i}\} \ \
{\boldsymbol {\tilde  \Lambda}}_{rd}={\rm{diag}}\{\lambda_{rd,i}\} \ \
{\boldsymbol  \Lambda}_{{\bf{\tilde F}}}={\rm{diag}}\{f_{i}\} \nonumber \\ &
{\boldsymbol  \Lambda}_{{\bf{\tilde P}}}={\rm{diag}}\{p_{i}\} \ \ {\boldsymbol \Lambda}_{\bf{W}}={\rm{diag}}\{w_i\}
\end{align}\end{small}the optimization problem (\ref{prob:1}) becomes
\begin{small}
\begin{align}
&\min_{f_i,p_i} \ \sum_{i=1}^N\frac{w_i(f_i^2\lambda_{rd,i}^2+p_i^2\lambda_{sr,i}^2+1)}
{(p_i^2\lambda_{sr,i}^2+1)(f_i^2\lambda_{rd,i}^2+1)} \nonumber \\
 &\ {\rm{s.t.}} \ \   \sum_{i=1}^N f_i^2 =P_r \ \  \sum_{i=1}^N p_i^2 =P_s,
\end{align}\end{small}where ${\boldsymbol { \tilde \Lambda}}_{a} ={\rm{diag}}\{\lambda_{a,i}\}$ means that the $i^{\rm{th}}$ diagonal element of the diagonal matrix ${\boldsymbol {\tilde  \Lambda}}_{a}$ is denoted as $\lambda_{a,i}$. This optimization is non-convex, and thus generally speaking it is difficult to solve. However, notice that when $p_i$'s are fixed, $f_i$'s can be computed as
\begin{small}
\begin{align}
\label{f_diag}
{f}_i=\left[ \left(\sqrt{\frac{w_i}{\mu_f\lambda_{rd,i}^2 }} \sqrt{\frac{p_i^2\lambda_{sr,i}^2}{1+p_i^2\lambda_{sr,i}^2}}
-\frac{1}{\lambda_{rd,i}^2}\right)^{+}\right]^{1/2}
,
\end{align}\end{small}where $\mu_f$ is the Lagrange multiplier which makes $\sum f_i^2=P_r$. On the other hand, when $f_i$'s are fixed, $p_i$'s can be computed as
\begin{small}
\begin{align}
\label{p_diag}
{p}_i=\left[ \left(\sqrt{\frac{w_i}{\mu_p\lambda_{sr,i}^2 }} \sqrt{\frac{f_i^2\lambda_{rd,i}^2}{1+f_i^2\lambda_{rd,i}^2}}
-\frac{1}{\lambda_{sr,i}^2}\right)^{+}\right]^{1/2},
\end{align}\end{small}where $\mu_p$ is the Lagrange multiplier which makes $\sum p_i^2=P_s$ hold. Notice that this iterative water-filling algorithm is guaranteed to converge, as discussed in \cite{Yu04}.

%The convergence of iterative water-filling solution is well-known.
%The performance of the proposed iterative algorithms definitely has a closed relationship with initial values. In order to overcome this weakness, we try to make the computation in the first iteration independent of the initial value. As (\ref{f_diag}) and (\ref{p_diag}) are symmetric, we take (\ref{f_diag}) for example. If in the first step, $f_i$'s is computed, the parts $\sqrt{{p_i^2\lambda_{sr,i}}/{(1+p_i^2\lambda_{sr,i})}}$'s rely on the $p_i$'s. It can be seen that in high SNR this part tends to be 1. Therefore, in the first iteration  we set $\sqrt{{p_i^2\lambda_{sr,i}}/{(1+p_i^2\lambda_{sr,i})}}=1$ to avoid initial value selection.
%Please notice that as SNR increases, this approximation becomes tighter and tighter. It means that in high SNR the proposed solution approaches the global optimally solution.

\noindent {\textbf{Special cases}}: Several existing algorithms can be considered as special cases of our proposed solution.

\noindent $\bullet$ When CSI is perfectly known, ${\bf{W}}={\bf{I}}$ and ${\bf{P}}={\bf{I}}$, the proposed solution for ${\bf{F}}$ reduces to that in \cite{Guan08}.

\noindent $\bullet$ When CSI is perfectly known and ${\bf{W}}={\bf{I}}$, the proposed solution for ${\bf{P}}$ and ${\bf{F}}$ reduces to that given in \cite{Rong09}.

\noindent $\bullet$ When the second hop channel is an identity matrix and noiseless, the proposed solution for source precoder design reduces to that given in \cite{Ding09}.

\section{Simulation Results and Discussions}

In this section, simulation results are presented to demonstrate the
performance of the proposed algorithm. For the purpose of
comparison, the algorithm based on the estimated channel only
(without taking the channel errors into account) \cite{Rong09} and the robust algorithm without source precoder design in \cite{Xing10} are also simulated. In the following, we consider an AF MIMO relay
system where the source, relay and destination are equipped with the
same number of antennas, i.e., $N_S=M_R=N_R=M_D=4$. The elements of
channel matrices ${\bf{H}}_{sr}$ and ${\bf{H}}_{rd}$ are randomly generated
as i.i.d. Gaussian distributed random variables.

 The widely used exponential correlation matrix ${\bf{R}}_{\alpha}=\{\alpha^{|i-j|}\}_{ij}$ is used to model the correlation matrix of ${\bf{D}}$, i.e., ${\bf{D}}{\bf{D}}^{\rm{H}}\propto {\bf{R}}_{\alpha}$. Then ${\boldsymbol \Psi}_{sr}={\boldsymbol \Sigma}_{rd}=({\bf{I}}_4+{\rm{SNR}}_{\rm{EST}}{\bf{R}}_{\alpha})^{-1}$ where ${\rm{SNR}}_{\rm{EST}}$ is the signal-to-noise ratio (SNR) in channel estimation process.
In the simulation, the weighting matrix is
selected as ${\bf{W}}={\rm{diag}}\{[0.3 \ 0.3 \ 0.2 \ 0.2]\}$. The noise covariance matrices are  ${\bf{R}}_{n_1}=\sigma_1^2{\bf{I}}_4$ and
${\bf{R}}_{n_2}=\sigma_{2}^2{\bf{I}}_4$. In data transmission stage the SNRs at relay and destination are defined as
$P_s/{\sigma_{1}}^2$ and $P_r/\sigma_2^2$, respectively.

\begin{figure}[!ht] \centering
\includegraphics[width=.4\textwidth]{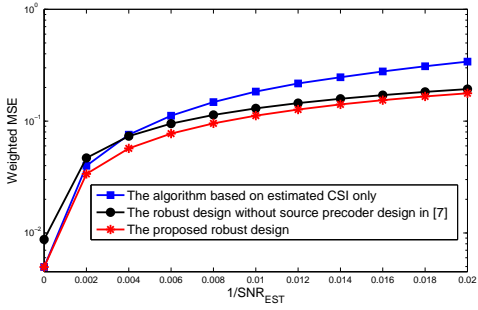}
\caption{Weighted MSEs of the detected data for different
algorithms, when $\alpha=0.3$ and
$P_s/\sigma_{1}^2=P_r/\sigma_2^2=30$dB.}\label{fig:2}
\end{figure}
For the source node, four independent data streams are transmitted and in each data stream, ${N_{Data}}=10000$ independent QPSK symbols
are transmitted. Each point in the figure is an average of 10000 independent channel realizations. Fig.~\ref{fig:2} shows the weighted MSEs at the destination for different algorithms when $\alpha=0.3$ and $P_s/\sigma_1^2=P_r/\sigma_2^2=30$dB. It can be seen that the performance of the proposed algorithm is always better than that based on the estimated CSI only. It can also be observed that the proposed  robust algorithm with source precoder design performs better than that without source precoder design in \cite{Xing10}, illustrating the importance of joint transceiver design involving source precoder.

\section{Conclusions}
Robust LMMSE transceiver design under Gaussian channel uncertainties for
dual-hop AF MIMO relay systems was investigated.  Exploiting channel estimation error statistics and using a general weighted MSE performance metric, the precoder matrix at the source, forwarding matrix at the relay and equalizer matrix at the destination were jointly optimized. It was found that several existing solutions are special cases of our proposed solution. The performance advantage of the proposed algorithm was demonstrated by the computer simulations.

\appendices

\section{}Based on (\ref{AMA}) and the fact that ${\boldsymbol{\tilde
\Lambda}}_{{\bf{A}}}$ and ${\boldsymbol{\tilde
\Lambda}}_{{\bf{M}}}$ are diagonal matrices with diagonal elements in decreasing order, the following identity holds
\begin{small}\begin{align}
\label{Apped_1}
{\bf{A}}^{\rm{H}}{\bf{M}}{\bf{A}}= {\bf{V}}_{\bf{A}}{\boldsymbol \Lambda}_1{\bf{V}}_{\bf{A}}^{\rm{H}},
\end{align}\end{small}where ${\boldsymbol{\Lambda}}_1={\boldsymbol{\tilde
\Lambda}}_{{\bf{A}}}{\boldsymbol{\tilde
\Lambda}}_{{\bf{M}}}{\boldsymbol{\tilde
\Lambda}}_{{\bf{A}}}$
is a diagonal matrix with diagonal elements in decreasing
order. Notice that (\ref{Apped_1}) in fact is an eigen-decomposition of ${\bf{A}}^{\rm{H}}{\bf{M}}{\bf{A}}$ with eigenvalues in decreasing order and ${\bf{V}}_{\bf{A}}$ is the corresponding unitary matrix.

Defining\begin{small}\begin{align}
\label{Apped_3}
{\bf{N}}&\triangleq( {\bf{\bar H}}_{rd}{\bf{\tilde
F}}{\boldsymbol{\Pi}}_{\bf{P}}^{-1/2}{\bf{K}}_1^{-1/2}{\bf{\bar
H}}_{sr}{\bf{P}})^{\rm{H}}({\bf{\bar H}}_{rd}{\bf{\tilde F}}{\bf{\tilde
F}}^{\rm{H}} {\bf{\bar H}}_{rd}^{\rm{H}}+{\bf{K}}_2)^{-1}\nonumber \\
& \times({\bf{\bar
H}}_{rd}{\bf{\tilde F}}{\boldsymbol{\Pi}}_{\bf{P}}^{-1/2}{\bf{K}}_1^{-1/2}{\bf{\bar
H}}_{sr}{\bf{P}}),
\end{align}\end{small}and together with the definitions of ${\bf{A}}$ and ${\bf{M}}$ in (\ref{A_M_Define}) the following equation holds
\begin{small}
\begin{align}
{\bf{A}}^{\rm{H}}{\bf{M}}{\bf{A}}={\bf{W}}^{1/2}{\bf{N}}{\bf{W}}^{1/2},
\end{align}
\end{small}based on which, the weighted MSE (\ref{weighted_MSE}) can be rewritten as \begin{small}
\begin{align}
{\rm{MSE}}_W({\bf{\tilde
 F}},{\bf{P}})&={\rm{Tr}}({\bf{W}})-
 {\rm{Tr}}({\bf{W}}{\bf{N}})\nonumber \\
 & \ge {\rm{Tr}}({\bf{W}})-{\sum}_i\lambda_i({{\bf{W}}})\lambda_i({\bf{N}}),
\end{align}\end{small}where $\lambda_i({\bf{Z}})$ denotes the $i^{\rm{th}}$ largest eigenvalue of ${\bf{Z}}$. Using Neumann inequality \cite{Marshall79}, for the minimum weighted MSE ${\bf{N}}$ and ${\bf{W}}$ have the same eigen-vectors. In other words, given the eigen-decomposition of ${\bf{W}}$ as
\begin{small}
\begin{align}
\label{Apped_W}
{\bf{W}}&={\bf{U}}_{\bf{W}}{\boldsymbol \Lambda}_{\bf{W}}{\bf{U}}_{\bf{W}}^{\rm{H}}
\end{align}
\end{small}where the diagonal elements of the diagonal matrix ${\boldsymbol \Lambda}_{\bf{W}}$ are in decreasing order, for the minimum MSE, ${\bf{N}}$ could be eigen-decomposed as
\begin{small}
\begin{align}
\label{Apped_N}
{\bf{N}}={\bf{U}}_{\bf{W}}{\boldsymbol \Lambda}_{\bf{N}}{\bf{U}}_{\bf{W}}^{\rm{H}}.
\end{align}\end{small}where ${\boldsymbol \Lambda}_{\bf{N}}$ is a diagonal matrix whose diagonal elements are in decreasing order. Based on (\ref{Apped_W}) and (\ref{Apped_N}), for the minimum MSE it holds that
 \begin{small}
\begin{align}
\label{Apped_2}
{\bf{A}}^{\rm{H}}{\bf{M}}{\bf{A}}={\bf{W}}^{1/2}{\bf{N}}{\bf{W}}^{1/2}={\bf{U}}_{\bf{W}}\underbrace{{\boldsymbol
\Lambda}_{\bf{W}}{\boldsymbol
\Lambda}_{\bf{N}}}_{={\boldsymbol \Lambda}_2}{\bf{U}}_{\bf{W}}^{\rm{H}}.
\end{align}\end{small}As the diagonal elements of the diagonal matrices ${\boldsymbol \Lambda}_{\bf{W}}$ and ${\boldsymbol \Lambda}_{\bf{N}}$ are positive and both in decreasing order, the diagonal elements of ${\boldsymbol \Lambda}_{2}$ are also in decreasing order. Clearly, the second equation of (\ref{Apped_2}) also denotes an eigen-decomposition of ${\bf{A}}^{\rm{H}}{\bf{M}}{\bf{A}}$. Comparing (\ref{Apped_1}) and (\ref{Apped_2}), it can be concluded that for the minimum MSE there exists an eigen-decomposition of ${\bf{A}}^{\rm{H}}{\bf{M}}{\bf{A}}$ with eigenvalues in decreasing order such that
\begin{small}
\begin{align}
 {\bf{V}}_{\bf{A}}={\bf{U}}_{\bf{W}}.
\end{align}
\end{small}

\end{document}